\documentclass[twocolumn,english,aps,prl,superscriptaddress]{revtex4-1}
\usepackage[T1]{fontenc}
\usepackage{color}
\usepackage{bm}
\usepackage{amsmath}
\usepackage{amssymb}
\usepackage{graphicx}

\makeatletter
\usepackage{bm}
\usepackage{graphicx}

\makeatother

\usepackage{babel}
\begin{document}

\title{Highly controllable and robust 2D Spin-Orbit Coupling for quantum gases}

\author{Wei Sun}
\affiliation{Shanghai Branch, National Research Center for Physical Sciences at Microscale
and Department of Modern Physics, University of Science and Technology
of China, Shanghai 201315, China}
\affiliation{Chinese Academy of Sciences Center for Excellence: Quantum Information and Quantum Physics,
University of Science and Technology of China, Hefei Anhui 230326,
China}

\author{Bao-Zong Wang}
\affiliation{Shanghai Branch, National Research Center for Physical Sciences at Microscale
and Department of Modern Physics, University of Science and Technology
of China, Shanghai 201315, China}
\affiliation{Chinese Academy of Sciences Center for Excellence: Quantum Information and Quantum Physics,
University of Science and Technology of China, Hefei Anhui 230326,
China}
\affiliation{International Center for Quantum Materials, School of Physics, Peking
University, Beijing 100871, China}
\affiliation{Collaborative Innovation Center of Quantum Matter, Beijing 100871,
China}

\author{Xiao-Tian Xu}
\author{Chang-Rui Yi}
\affiliation{Shanghai Branch, National Research Center for Physical Sciences at Microscale
and Department of Modern Physics, University of Science and Technology
of China, Shanghai 201315, China}
\affiliation{Chinese Academy of Sciences Center for Excellence: Quantum Information and Quantum Physics,
University of Science and Technology of China, Hefei Anhui 230326, China}

\author{Long Zhang}
\affiliation{International Center for Quantum Materials, School of Physics, Peking
University, Beijing 100871, China}
\affiliation{Collaborative Innovation Center of Quantum Matter, Beijing 100871, China}

\author{Zhan Wu}
\author{Youjin Deng}
\affiliation{Shanghai Branch, National Research Center for Physical Sciences at Microscale
and Department of Modern Physics, University of Science and Technology
of China, Shanghai 201315, China}
\affiliation{Chinese Academy of Sciences Center for Excellence: Quantum Information and Quantum Physics,
University of Science and Technology of China, Hefei Anhui 230326, China}

\author{Xiong-Jun Liu}
\email{xiongjunliu@pku.edu.cn}
\affiliation{International Center for Quantum Materials, School of Physics, Peking
University, Beijing 100871, China}
\affiliation{Collaborative Innovation Center of Quantum Matter, Beijing 100871,
China}

\author{Shuai Chen}
\email{shuai@ustc.edu.cn}
\author{Jian-Wei Pan}
\email{pan@ustc.edu.cn}

\affiliation{Shanghai Branch, National Research Center for Physical Sciences at Microscale
and Department of Modern Physics, University of Science and Technology
of China, Shanghai 201315, China}
\affiliation{Chinese Academy of Sciences Center for Excellence: Quantum Information and Quantum Physics,
University of Science and Technology of China, Hefei Anhui 230326,
China}

\begin{abstract}
We report the realization of a robust and highly controllable two-dimensional (2D) spin-orbit (SO) coupling with topological non-trivial band structure. By applying a retro-reflected 2D optical lattice, phase tunable Raman couplings are formed into the anti-symmetric Raman lattice structure, 
and generate the 2D SO coupling with precise inversion and $C_4$ symmetries, leading to considerably enlarged topological regions.
The life time of the 2D SO coupled Bose-Einstein condensate reaches several seconds, which enables the exploring of fine tuning interaction effects.
These essential advantages of the present new realization open the door to explore exotic quantum many-body effects and non-equilibrium dynamics with novel topology.
\end{abstract}
\maketitle

%\section{introduction}

Spin-orbit (SO) coupling, as a fundamental quantum effect, is an essential ingredient in many topological phases of 
quantum matter, including the topological insulators~\cite{TI1,TI2} and topological semimetals~\cite{semimetal1,semimetal2,xu2015discovery,lv2015experimental}.
In the field of ultracold atoms which can offer extremely clean and controllable platforms for quantum simulation,
synthetic SO coupling schemes were proposed~\cite{Ruseckas2005,Osterloh2005,Liu2005,Tudor2007,Liu2009} via light-atom interactions
which flip atomic spins and transfer momentum simultaneously. 
The 1D SO coupling has now been realized routinely in the experiments for both ultracold bosons~\cite{key-6,key-7} and fermions~\cite{key-8,key-9} as continuum atom gases or trapped in optical lattices~\cite{Engels2013,Engels2014}.
Great interests have been drawn in emulating SO effects~\cite{review1,review2,review3,review4,review5,YeLiu1,YeLiu2} and topological phases with ultracold atoms~\cite{LiuSHE,ZhuSHE,Zhang2008,Sato2009,LiuFSHE2009,Goldman2010,Zhu2011PRL,Qu2013,Zhang2013,Liu2014PRL-a,Xu2015PRL,Chan2017PRL,Ueda2017PRL,Chan2017PRL-b}. In particular, the simulation of broad classes of topological quantum states or phase transitions with quantum gases necessitates to 
synthesize 2D or higher dimensional SO couplings.

The 2D SO couplings with nontrivial band topology were recently realized for $^{87}$Rb Bose-Einstein condensate (BEC)~\cite{wu2016realization} 
and $^{40}$K Fermi gases~\cite{huang2016experimental,meng2016experimental}.  
However, they suffer from the instability of optical lattices with long loop lines~\cite{wu2016realization,Liu2014PRL-a} or heating with complex Raman couplings between multiple ground states~\cite{huang2016experimental,meng2016experimental}. 
These shortcomings make it hard to explore the exotic topological quantum phases with high precision.  
Realization of 2D SO coupled atom gas with high controllability and long lifetime, which are essential 
for quantum simulations of exotic states, are highly desired. 

In this Letter, we report on the synthesis of a 2D SO coupling with high controllability, stability, and long lifetime.
The realization is based on a new optical Raman lattice scheme~\cite{Baozong2017}, with the optical instability and heating %all the shortcomings 
in the previous works~\cite{wu2016realization,huang2016experimental,meng2016experimental} all being overcome.
The 2D SO coupling exhibits precise inversion and $C_4$ symmetries, which are the key features and 
bring important new physics, including a highly resolved crossover between 1D and 2D SO couplings,
a long lifetime of the 2D SO coupled BEC up to several seconds, one order of magnitude longer than those in 
Refs.~\cite{wu2016realization,huang2016experimental,meng2016experimental}, the detection with high precision of the critical transition point from stripe phase to plane-wave phases~\cite{key-6, key-14}, and the considerably broadened topological regions. This work paves the way for further studies of quantum many-body physics and quantum non-equilibrium dynamics with novel topology.

%\section{The scheme}
\begin{figure}
\includegraphics[scale=0.8]{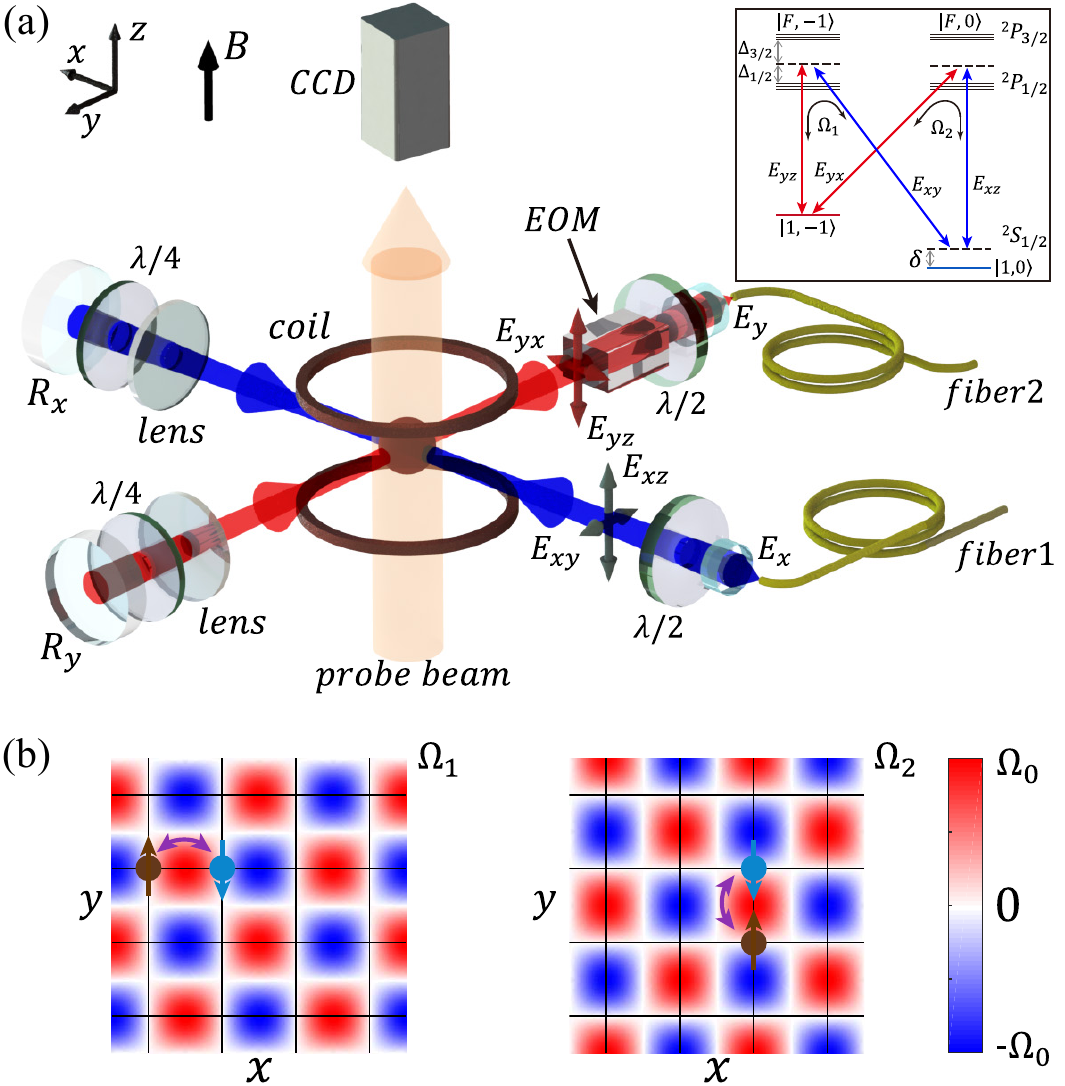}
\caption{\label{realization}
(a) Experimental setup. %of 2D SO coupling with retro-reflected optical lattices.
The $\bold B$ field along z-axis generates the Zeeman splitting and the quantization axis of the atoms.
The red and blue lines are lasers to construct the 2D lattice and Raman couplings.
The $\lambda/2$-waveplates are used to generate two orthogonal polarization components of the lattices.
The $\lambda/4$-waveplates are phase retarders, which are applied to form the anti-symmetric Raman coupling lattices.
The EOM is applied to tune the relative phase $\delta\varphi$ between two Raman couplings $\Omega_{1}$ and $\Omega_{2}$.
insets: level structure and Raman coupling scheme.
(b) The anti-symmetric structure of the two Raman couplings $\Omega_1$ and $\Omega_2$ in real space. The grid represent the square optical lattice $V_{\text{latt}}(x,y)$.}
\end{figure}

The new scheme of the 2D SO coupling~\cite{Baozong2017} with ultracold $^{87}$Rb atoms is illustrated in Fig.\ref{realization}(a).
The 2D square lattice $V_{{\rm {latt}}}(x,y)$ is constructed with two linear polarized laser beams $\boldsymbol{E}_{x}$
and $\boldsymbol{E}_{y}$ (the blue and red lines in Fig.~1(a)),
with wavelength of $\lambda=787\text{nm}$ (wave vector $k_0=2\pi/\lambda$ and recoil energy $E_{\text{r}}=\hbar^{2}k_{0}^{2}/2m$).
The beams pass through $\lambda/2$ waveplates to generate two orthogonally polarized components $E_{xz}$ ($E_{yz}$) and $E_{xy}$ ($E_{yx}$) before shining on the atoms. 
The key setting is that they are retro-reflected back by two mirrors ($R_x$ and $R_y$) to form a 2D square lattice, 
where two $\lambda/4$ wave-plates with optical axis along $\hat{z}$ direction are placed in front of the mirrors %$R_x$ and $R_y$
as the phase retarders to generate $\pi$ phase delay between the two orthogonal polarization components.
The 2D lattice potential then reads
\begin{equation}
V_{{\rm {latt}}}(x,y)=V_{0x}\cos^{2}k_{0}x+V_{0y}\cos^{2}k_{0}y,
\end{equation}
with lattice depth $V_{0x(y)}\propto |E_{xy(yx)}|^{2}-|E_{xz(yz)}|^{2}$~\cite{Baozong2017,key-24}.
We tune $V_{0x}=V_{0y}=V_{0}$ to have a symmetric 2D lattice.

The spin up/down states are defined via two magnetic sub-levels in $F=1$ manifold, i.e.~$\left|1,-1\right\rangle$ and $\left|1,0\right\rangle$. 
A bias magnetic field $\boldsymbol{B}$ of $14.5{\rm {G}}$ is applied in $\hat{z}$-direction, giving a Zeeman split of $10.2{\text {MHz}}$.
The $|1,1\rangle$ state is effectively suppressed due to the large quadratic Zeeman shift ($\epsilon\approx8.2E_{{\rm {r}}}$). 
The Raman couplings are generated by the orthogonal polarization pairs ($E_{xz}, E_{yx}$) and ($E_{xy}, E_{yz}$), respectively, realizing the double-$\Lambda$ configuration, 
with $\Omega_{1}=\Omega_{01}\sin{k_{0}x}\cos{k_{0}y}$ and $\Omega_{2}=\Omega_{02}\cos{k_{0}x}\sin{k_{0}y}$,  shown in Fig.~1(b). 
The coupling strength $\Omega_{01}\propto|E_{yz}||E_{xy}|$ and $\Omega_{02}\propto|E_{yx}||E_{xz}|$~\cite{Baozong2017,key-24}.
Due to the anti-symmetric lattice structure, the onsite Raman coupling vanishes. 
Only in the tunneling, the atoms experience non-zero coupling strength which causes the spin-flip Raman diffraction.
This is essential for the generation of non-trivial topological band structure in the 2D SO coupling system~\cite{wu2016realization,Baozong2017}.
An electro-optical phase modulator (EOM) is placed in one of the beam path to tune the relative phase $\delta \varphi$ between $\Omega_{1}$ and $\Omega_{2}$. 
The total Raman coupling term reads
%\begin{equation}
%\Omega_{R}(x,y)=\Omega_{1}\sigma_x \pm \Omega_{2}\sigma_y .
%\end{equation}
\begin{equation}
\Omega_{R}(x,y)=\left( \begin{array}{cc}
0 & \Omega_1 + e^{i\delta\varphi} \Omega_2 \\
\Omega_1^\ast + e^{-i\delta\varphi}\Omega_2^\ast & 0
\end{array} 
\right)
\end{equation}
We adjust the two $\lambda/2$ waveplates to set $\Omega_{01}=\Omega_{02}=\Omega_{0}$ so that 
$\Omega_{R}(x,y)$ satisfies reflective anti-symmetry. 
An optimal 2D SO coupling is achieved with $\delta\varphi = \pm\pi/2 $. % under reflections of the x- and y-coordinate.}% 

In this scheme, the initial and propagation phases of the laser beams are static global phases and can be neglected~\cite{Baozong2017,key-24}.
The total Hamiltonian reads
\begin{equation}
\hat{H}=\frac{\bm p^2}{2m}+V_{\rm latt}(x,y)+\Omega_{R}(x,y)+\frac{\delta}{2}\sigma_z,
\end{equation}
where $m$ is the atomic mass, $\delta$ is the two-photon Raman detuning. 
It is the realization of a minimal quantum anomalous Hall (QAH) model driven by the SO coupling~\cite{Liu2014PRL-a}.
Different from the previous realization of 2D SO coupling~\cite{wu2016realization}, this Hamiltonian has the precise inversion and $C_4$ symmetries, which lead to nontrivial physics as presented below.

%\section{Crossover of 1D-2D SO couplings with high resolution}
\begin{figure}
\includegraphics[scale=0.82]{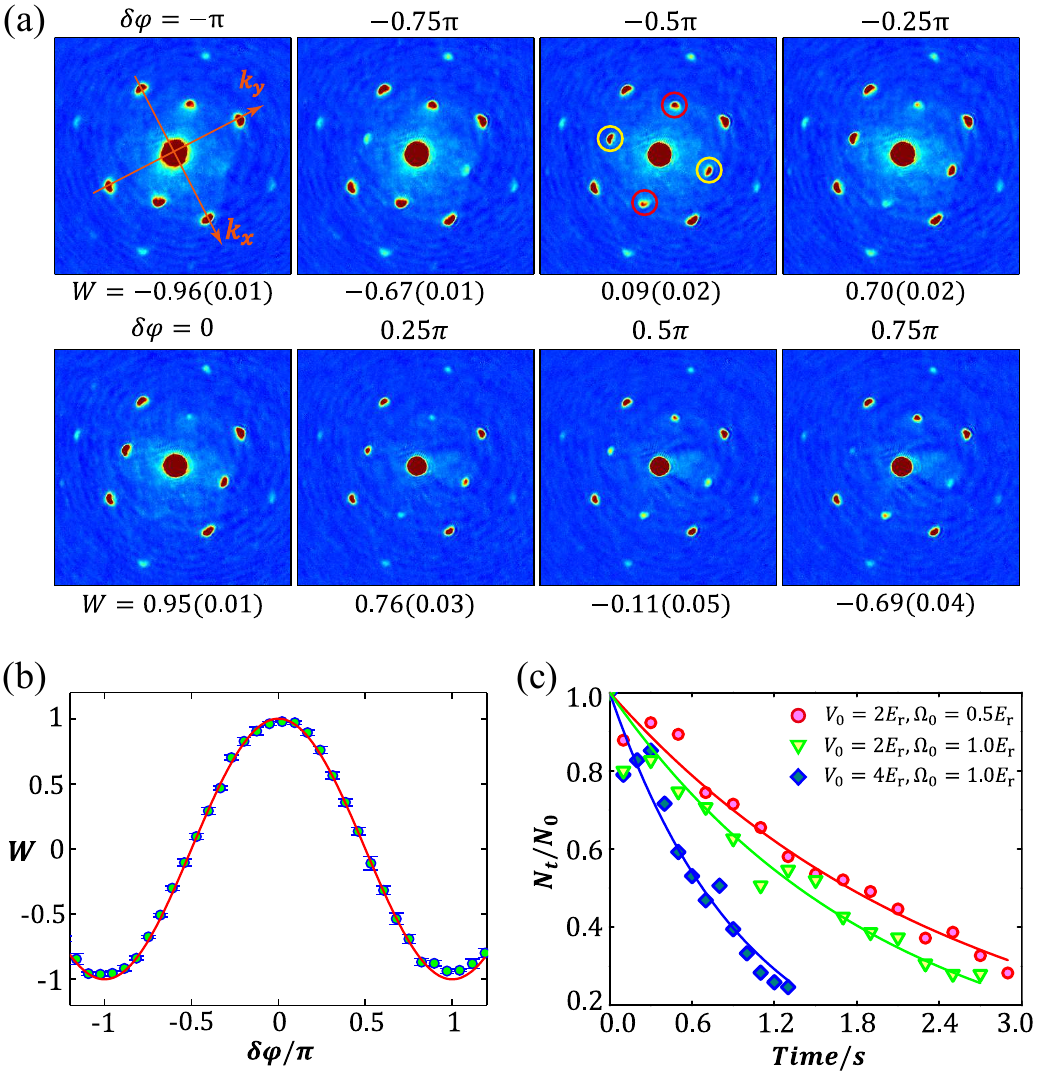}
\caption{\label{ground_state}
(a) TOF image of the SO coupled BEC with various phases $\delta\varphi$.
The images with $\delta\varphi=-\pi,0$ correspond to 1D SO coupling, and those with $\delta\varphi=\pm\pi/2$ are for symmetric 2D SO coupling.
(b) $W$ versus $\delta\varphi$. The blue circles are experimental data, with the error bar for the standard statistical error. 
The red solid line is the theoretical curve without fitting.
(c) The measured life time $\tau$ of 2D SO coupled BEC with $\delta\varphi=\pi/2$.
By fitting with an exponential decay, we obtain the lifetime as $\tau=2.52\pm0.07\text{s}$ (red line), $2.00\pm0.09\text{s}$ (green line) and $0.980\pm0.05\text{s}$ (blue line). }
\end{figure}

{\it Properties of 2D SO coupled BEC.}
The BEC with $3.0\times10^5$ $^{87}\text{Rb}$ atoms is prepared in $|1,-1\rangle=|\left\uparrow\right\rangle$ state.
The 2D lattice and Raman coupling beams %($\bm{E}_{x}$ and $\bm{E}_{y}$)
are ramped up adiabatically in $80\text{ms}$, and the BEC is loaded into the ground state of the Hamiltonian (Eq.~3) with $V_0=4.0E_{\rm r}$, $\Omega_{0}=1.0E_{\rm r}$ and $\delta=0$. 
By manipulating the voltage on the EOM, we can tune $\delta\varphi$ continuously over $2\pi$ range, which governs the interference of the Raman couplings $\Omega_{1,2}$~\cite{key-24} and leads to the crossover between 1D and 2D SO couplings. 
For detection, a time-of-flight (TOF) image is taken after the BEC is free released for $25\text{ms}$. 

A highly resolved crossover between 1D and 2D SO couplings is observed in Fig.~\ref{ground_state}(a,b). 
The major atom cloud in $\left|\uparrow\right\rangle$ state stays at momentum $(k_{x},k_{y})=(0,0)$.
The four atom clouds with momenta $\bold k=(\pm k_{0},\pm k_{0})$ in diagonal and off-diagonal directions are in $\left|\downarrow\right\rangle$ state diffracted by the Raman coupling term.  %spin-down state.
As shown in Fig.\ref{ground_state}(a), the distribution of the four $\left|\downarrow\right\rangle$ atom clouds varies versus $\delta\varphi$, which characterize the interference of the Raman coupling lattices $\Omega_{1}$ and $\Omega_{2}$.
We define $W(\delta\varphi)$ to quantify this interference by
\begin{eqnarray}
W(\delta\varphi)=\frac{N_{\hat{x}+\hat{y}}-N_{\hat{x}-\hat{y}}}{N_{\hat{x}+\hat{y}}+N_{\hat{x}-\hat{y}}},
\end{eqnarray}
where $N_{\hat{x}+\hat{y}}$ ($N_{\hat{x}-\hat{y}}$) is the total number of atoms
in the diagonal (off-diagonal) direction, 
as denoted by red (yellow) circles in Fig.~\ref{ground_state}(a), 
For the 1D SO coupling with $\delta\varphi=\pi$ ($0$), 
the Raman diffracted atoms are only in the diagonal (off-diagonal) direction, and 
we have $W=-0.96\pm0.01$ ($0.95\pm0.01$).
For the optimal 2D SO coupling with $\delta\varphi=\pm\pi/2$, 
the Raman diffracted atoms are almost evenly distributed, and
we obtain $W=-0.11\pm0.05$ and $0.09\pm0.02$.
%giving the optimal 2D SO coupling with $C_4$ symmetry. 
In comparison with Ref.~\cite{wu2016realization}, the quality of Fig.~\ref{ground_state}(a,b)
is significantly improved.

Further, the lifetime of the BEC with 2D SO coupling ($\delta\varphi=\pi/2$) is measured. 
After adiabatically loading the condensate, 
we measure the normalized number of condensate atoms $N(t)/N(0)$ as a function of the holding time $t$.
The results, shown in Fig.~\ref{ground_state}(c) for different lattice depths $V_0$ and Raman coupling strength $\Omega_{0}$, 
are fitted to an exponential function $e^{-t/\tau}$.
We obtain the lifetime $\tau=0.980\pm0.05\text{s}$ for $(V_0, \Omega_0) = (4.0, 1.0)E_{\text{r}}$, 
 $2.00\pm0.09\text{s}$ for $(4.0,2.0)E_{\text{r}}$,   %$V_{0}=4.0E_{\text{r}}$ ($2.0E_{\text{r}}$), $\Omega_{0}=1.0E_{\text{r}}$ 
and $2.52\pm0.07\text{s}$ for $(2.0,0.5)E_{\text{r}}$.  %  $V_{0}=2.0E_{\text{r}}$ and $\Omega_{0}=0.5E_{\text{r}}$.
Compared to Refs.~\cite{wu2016realization,huang2016experimental,meng2016experimental}, 
the lifetime of quantum gas is enhanced over one order of magnitude.

{\it Ground state phase transition.}
The high controllability, stability, and especially the long lifetime of the present 2D SO coupling, make it feasible to study the ground state phase transition driven by interaction, for which the longtime relaxation is essential for the system to reach the equilibrium~\cite{key-6,key-14}.
In particular, we measure the stripe phase (ST) and magnetic phase (MG) of the ground state for the present 2D SO coupled BEC (with $\delta\varphi=\pi/2$)~\cite{Pan2016}.
To observe and distinguish the two phases, the magnetization $M_0$ is defined as~\cite{key-14}, %and is quantified as
\begin{equation}
M_{0}=\frac{N_{0\uparrow}-N_{0\downarrow}}{N_{0\uparrow}+N_{0\downarrow}},
\end{equation}
where $N_{0\uparrow}$ ($N_{0\downarrow}$) represents the number of condensate atoms in the $\left|\uparrow\right\rangle$ ($\left|\downarrow\right\rangle$) state. 
In this measurement, $10^{6}$ atoms right above the BEC critical temperature, 
with half-and-half spin states, are loaded adiabatically into the 2D SO coupled system 
with fixed ($V_{0}=2.0E_{\text{r}}$,  $\delta=0$) and variable $\Omega_{0}$. 
The atoms are further cooled to pure condensate in $800\text{ms}$ and held for another $1.5\text{s}$ for relaxing to equilibrium before detection. Finally, the spin-resolved TOF image is taken to measure $N_{0\uparrow}$ and $N_{0\downarrow}$.

\begin{figure}[h]
\includegraphics[scale=0.4]{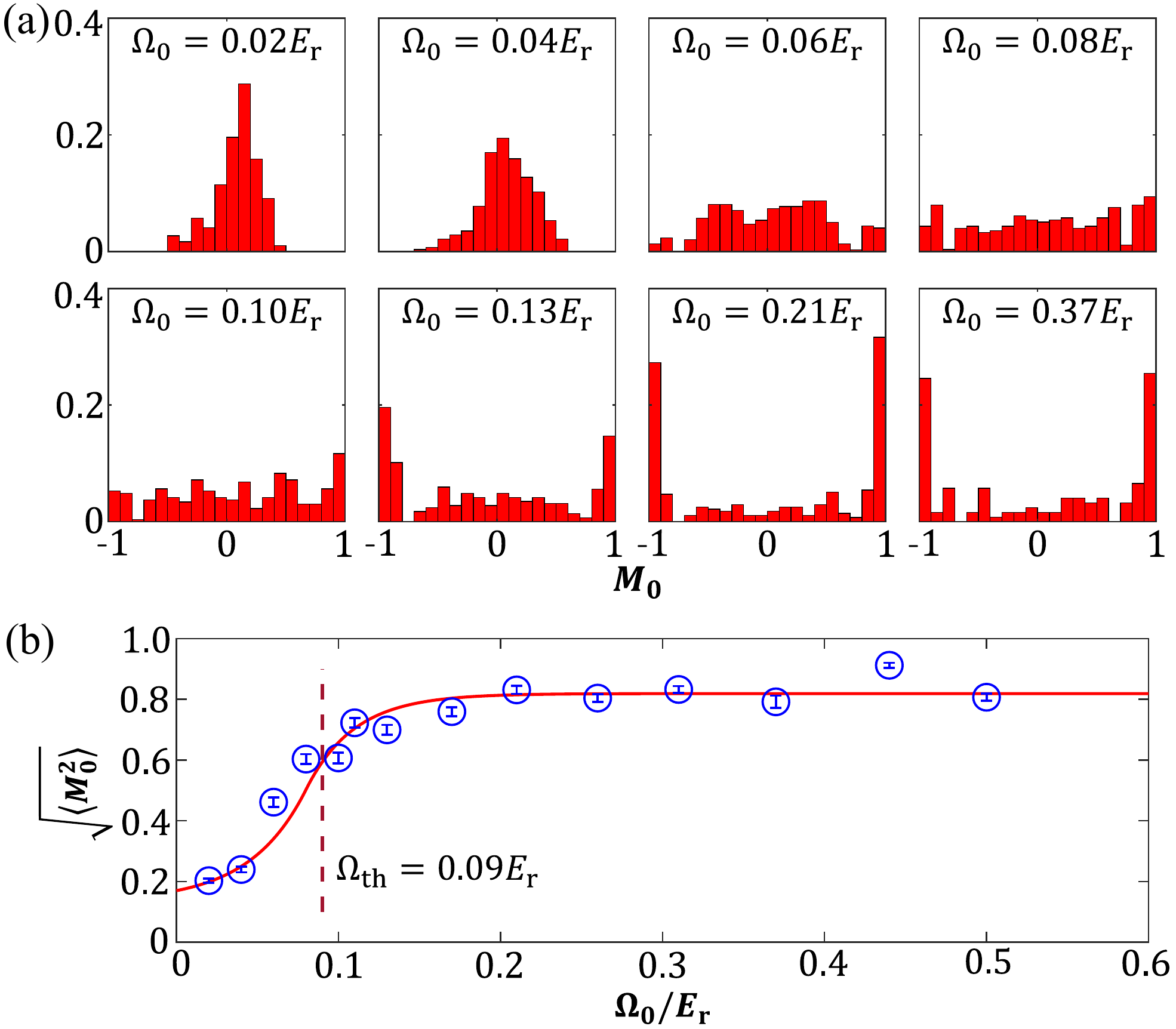}
        \caption{\label{STMG}Phase transition of the 2D SO coupled BEC between the ST and the MG phase. (a) Magnetization histograms with varying $\Omega_{0}$ for nearly pure condensate. (b) $\sqrt{\left \langle M_{0}^2\right\rangle}$ as a function of $\Omega_{0}$. The error bars give the standard statistical errors. %transferred from the measurement.
The red solid line is an arbitrary fit. The dashed line shows the theoretical calculation of phase transition point.}
\end{figure}

We repeat the measurement hundreds of times at every $\Omega_{0}$. 
Each value of $M_{0}$ is recorded to obtain the histograms for statistical analysis as shown in Fig.~\ref{STMG}(a). 
For very small $\Omega_{0}$ (as $0.04E_{\text{r}}$ and $0.02E_{\text{r}}$), the histogram of the magnetization shows a single peak centered at $M_{0}\approx0$, 
corresponding to the ST phase. 
For larger $\Omega_{0}$ (as $0.21E_{\text{r}}$ and $0.37E_{\text{r}}$), the histogram of $M_0$ shows two sharp peaks close to $\pm 1$, giving the MG phase~\cite{key-14}. 
In contrast to the ideal thermodynamic case which would exhibit an infinitely sharp transition between the two- and single-peak structures, 
the histogram broadens and flattens gradually in the moderate regime (from $\Omega_{0}=0.06E_{\text{r}}$ to $0.10E_{\text{r}}$).
This is due to the very small energy difference between the ST and MG states near the phase transition point, 
which leads to very long relaxation time to reach the equilibrium and domains with various sizes may form during the relaxation 
(see Supplementary Materials (SM) for details, which include Ref.~\cite{Amara1993}). 

The phase transition is determined from the turning point in the plot of $\sqrt{\left \langle M_{0}^2\right\rangle}$ versus
$\Omega_{0}$ (Fig.~\ref{STMG}(b)). %, {\color{blue} with $\sqrt{\left \langle M_{0}^{2} \right \rangle} \approx 0$ (1) for the ST (MG) phase.}
% and $\sqrt{\left \langle M_{0}^{2} \right \rangle}\approx1$ means MG phase. 
 The fit gives the critical point as $\Omega_{\text{c}}=0.08\pm0.02E_{\text{r}}$, 
 consistent with the theoretical prediction $\Omega_{\text{th}}=0.09E_{\text{r}}$~\cite{Pan2016}.
The experimental determination of $\Omega_{\text{c}}$, which is difficult in Ref.~\cite{wu2016realization}, 
shows the feasibility of the present realization for exploring many-body effects.

{\it Extended topological region.}
We proceed to measure the band topology, with $\delta\varphi=\pi/2$~\cite{Baozong2017}.
The topology of the $s$-band can be determined directly by the product of the signs of the spin-polarizations %$P_{0}\text{(\ensuremath{\boldsymbol{q}})}$ 
at the four symmetric momenta $\{\Lambda_{j}\}=\{\Gamma(0,0),X_{1}(0,\pi),X_{2}(\pi,0),M(\pi,\pi)\}$ in the FBZ~\cite{key-26}.
At low temperature regime, only the $s$-band is effectively populated. The measured spin-polarization $P\text{(\ensuremath{\boldsymbol{q}})}$ is dominated by the spin-polarization in the lowest $s$-band $P_{0}\text{(\ensuremath{\boldsymbol{q}})}$. Then we have
\begin{eqnarray}
\Theta=\Pi_{j=1}^{4}{\rm sgn}[P(\Lambda_{j})],
\end{eqnarray}
with the topologically trivial and non-trivial phases corresponding to $\Theta=1$
and $-1$, respectively~\cite{key-26}.
The Chern number of the lowest band is then obtained as
\begin{eqnarray}
\mathcal C_{1}^{(0)}=-\frac{1-\Theta}{4}\sum_{j=1}^{4}{\rm sgn}[P(\Lambda_{j})].
\end{eqnarray}

\begin{figure}[t]
\includegraphics[scale=0.8]{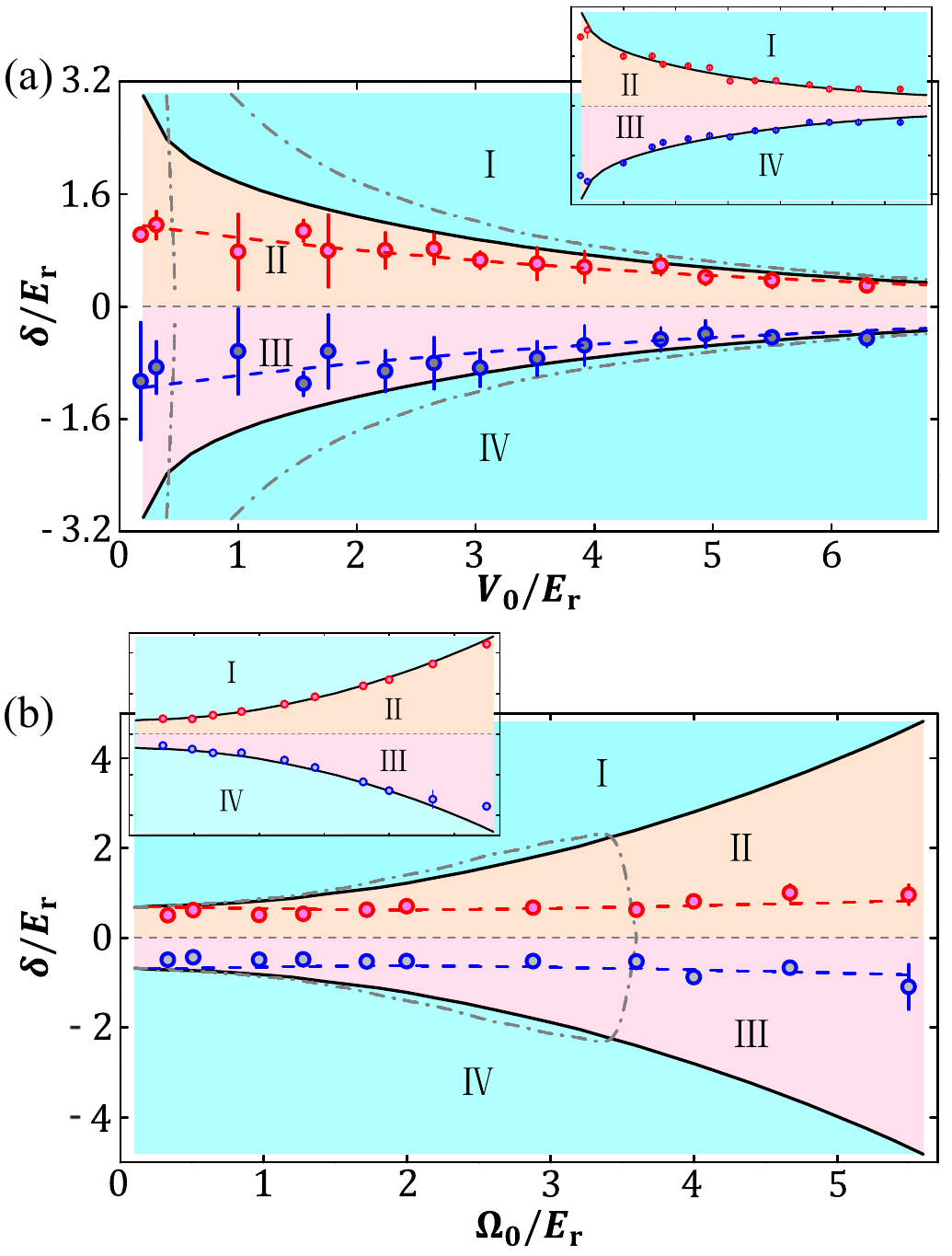}
\caption{\label{phase_diagram}Topological phase diagram of the lowest band.
The regions I and IV are trivial with the Chern number $\mathcal C_1^{(0)}=0$.
The areas II and III are topological non-trivial with $\mathcal C_1^{(0)}=\pm1$, respectively.
The black solid lines are the topological phase boundaries in the present scheme,
and the gray dash-dot lines are that of %the topological phase boundaries with
asymmetric Raman coupling in Ref.~\cite{wu2016realization}.
The circles with statistical error bars are the phase boundaries measured in the experiment. 
(a) $\delta-V_0$ plane with $\Omega_0=0.5E_{\rm{r}}$. %the numerical calculation and measurements consist. 
(b) $\delta-\Omega_{0}$ plane with $V_0=4.0E_{\rm{r}}$.
The measurements deviate from the calculation at small $V_0$ and large $\Omega_0$, due to the higher band and heating effects.
Insets: by considering the thermal effect and removing the high-band contribution from the experimental data, the phase boundaries are found to be fit to the theoretical calculation.}
\end{figure}

In the experiment, about $2.0\times10^5$ thermal atoms at temperature $100\text{nK}$ are filled into the lowest band, with the higher band population being relatively small. The spin-resolved TOF imaging measures the atom densities at spin-up ($n_{\uparrow}(\boldsymbol{q})$) and spin-down ($n_{\downarrow}(\boldsymbol{q})$) in momentum space.
The spin-polarization at quasi-momentum $\bm q$ is obtained as $P\text{(\ensuremath{\boldsymbol{q}})}=\text{[\ensuremath{n_{\uparrow}(\boldsymbol{q})} -\ensuremath{n_{\downarrow}\text{(\ensuremath{\boldsymbol{q}})}}]}/[\ensuremath{n_{\uparrow}(\boldsymbol{q})} +\ensuremath{n_{\downarrow}\text{(\ensuremath{\boldsymbol{q}})}}]$.
By varying $\delta$, we take each of the $P(\Lambda_{j})$ to calculate $\Theta(\delta)$ and 
obtain the topological phase boundary (see the SM for details).
The experimental results (circles) and numerical simulation (solid lines) are shown in Fig.~\ref{phase_diagram}, 
for the $\delta$-$V_0$ plane with fixed $\Omega_{0}=0.5E_{\text{r}}$ [for (a)] and 
for the $\delta$-$\Omega_{0}$ plane with fixed $V_0=4.0E_{\text{r}}$ [for (b)].
The regions II and III (I and IV) are for topological (trivial) bands with $\mathcal C_1^{(0)}=\pm1$ ($\mathcal C_1^{(0)}=0$).
The Chern number changes sign when $\delta$ crosses zero.  
The topological region extends to the limit $V_0\rightarrow0$ (a) and $\Omega_0\rightarrow\infty$ (b),
as predicted in Ref.~\cite{Baozong2017}.
This is in sharp contrast to the case without $C_4$ symmetry~\cite{wu2016realization,
where the topological phases (dash-dot lines) end at}  $V_{0}^c\approx0.46E_{\text{r}}$ and $\Omega_0^c\approx3.6E_{\text{r}}$, respectively. 
The broad topological region with weak lattice limit $V_0 \rightarrow 0$ can enable to explore topological physics in nearly continuous regimes.

The directly measured phase boundaries in Fig.~\ref{phase_diagram} are narrower than those predicted in theory for the following two reasons.
First, as  $V_0$  decreases and/or  $\Omega_0$ increases, the system gradually deviates from the tight-binding condition. 
The high-band effect suppresses the net magnitudes of spin-polarizations $P(\Lambda_j)$ at symmetric momenta. 
Second, the heating becomes serious for large $\Omega_0$ (e.g. for $\Omega_0=4.65E_{\rm{r}}$, the temperature of the atoms increases from $100\text{nK}$ to nearly $300\text{nK}$). 
These effects reduce the resolution of $s$-band spin-polarization due to the high-band population. Subtracting such high-band contribution from the measured total spin-polarization we can obtain the spin-polarization contributed only from the lowest-band states (see the SM for details). 
We then recalculate the topological index and find that the results agree well with the numerical solution (the insets of Fig.\ref{phase_diagram}).

In conclusion, we have synthesized a highly controllable and robust 2D SO coupling % topological Bose gas 
for quantum gases via an anti-symmetric Raman lattice scheme. It exhibits precise inversion and $C_4$ symmetries.
The lifetime of the 2D SO coupled BEC reaches several seconds, 
which enable us to study the delicate interaction-driven ground state phase transition where long time relaxation is required. 
Further, the precise symmetries qualitatively enlarge the topological region in the phase diagram, and leads to topological bands with arbitrary Raman coupling strength and lattice depth. 
Moreover, the present scheme can be also applied to other Bosonic or Fermionic quantum gases systems.
As for  $^{40}\text{K}$, the estimated heating is about 5 times larger than that of $^{87}\text{Rb}$,
giving a life time of hundreds of milliseconds (see SM which include Ref.~\cite{dipoletrap}).  It is thus expected that rich quantum many-body phenomena with non-trivial topology can be also studied in 2D SO coupled fermionic systems.   
The high controllability and stability of the 2D SO coupling open intriguing opportunities to explore novel topological quantum effects and far-from-equilibrium quantum dynamics with topology.

We thank J. -S Pan and W. Yi for the helpful discussion. This work was supported by the National Key R\&D Program of China (under grants 2016YFA0301601 and 2016YFA0301604), National Nature Science Foundation of China (under grants N0. 11674301, No. 11574008, No. 11761161003, and No. 11625522), and the Thousand-Young-Talent Program of China.
%%%%%%%%%%%%%%%%%%%%%%%%%%%%%%%%%%%%%%%%%%%%%%%%%%%%%%%%%%%%%%%%%%%%%%%%%%%%

\end{document}